\pdfoutput=1
\documentclass[12pt]{article}
\usepackage{placeins}
\usepackage{mathtools}
\usepackage[numbers,sort&compress]{natbib}
\usepackage{hyperref}
\usepackage{graphicx}
\usepackage{booktabs}
\usepackage{array}
\usepackage{longtable}
\usepackage{geometry}
\usepackage{amsmath,amssymb,amsfonts}
\usepackage{xcolor}
\usepackage{etoolbox}
\usepackage{adjustbox}
\AtBeginEnvironment{table}{\small}
\usepackage[font=small,labelfont=bf]{caption}
\newcommand{\dd}{\mathrm{d}}
\newcommand{\pp}{\partial}
\newcommand{\Bo}{\mathit{Bo}}
\newcommand{\Ca}{\mathit{Ca}}

\newcommand{\tp}{^{\mathsf T}}

\newcommand{\vect}[1]{\boldsymbol{#1}}
\newcommand{\mat}[1]{\mathbf{#1}}
\newcommand{\q}{\vect{q}}
\newcommand{\qd}{\dot{\vect{q}}}
\newcommand{\qdd}{\ddot{\vect{q}}}
\newcommand{\p}{\vect{p}}
\newcommand{\x}{\vect{x}}
\newcommand{\e}{\vect{e}}
\newcommand{\bb}{\vect{b}}
\newcommand{\f}{\vect{f}}
\newcommand{\vv}{\vect{v}}
\newcommand{\M}{\mat{M}}
\newcommand{\G}{\mat{G}}
\newcommand{\Lc}{\mat{L}}
\newcommand{\A}{\mat{A}}
\newcommand{\T}{\mat{T}}
\newcommand{\K}{\mat{K}}
\newcommand{\D}{\mat{D}}
\newcommand{\C}{\mat{C}}

\newcommand{\Pim}{\mat{\Pi}}
\newcommand{\I}{\mat{I}}
\newcommand{\Jm}{\mat{J}}
\newcommand{\Rm}{\mat{R}}
\newcommand{\Bm}{\mat{B}}
\newcommand{\sgn}{\operatorname{sgn}}
\newcommand{\Sgn}{\operatorname{Sgn}}
\newcommand{\Ham}{\mathcal{H}}
\newcommand{\Dis}{\mathcal{D}}
\newcommand{\Gop}{\mathcal{G}}
\newcommand{\Jop}{\mathcal{J}}

\graphicspath{{figs/}}

\title{A curvature-corrected port-Hamiltonian reduced-order model for dissipative sloshing with contact-line hysteresis in low-gravity tanks}
\author{%
\parbox{\dimexpr\textwidth-2\tabcolsep\relax}{\centering
Gianni Cassoni\textsuperscript{1}\,\thanks{Corresponding author: \texttt{gianni.cassoni@uniroma3.it}}, Fabrizio Gennari\textsuperscript{2}, Jacopo Serafini\textsuperscript{1}, Pierangelo Masarati\textsuperscript{3}\\[4pt]
\small \textsuperscript{1}Department of Civil, Computer Science and Aeronautical Technology Engineering, Roma Tre University, Via Vito Volterra 62, 00146 Rome, Italy\\
\small \textsuperscript{2}Thales Alenia Space Italia, Via Saccomuro 24, 00131 Rome, Italy\\
\small \textsuperscript{3}Department of Aerospace Science and Technology, Politecnico di Milano, Via La Masa 34, 20156 Milan, Italy}}
\date{}

\begin{document}
\maketitle
\begin{abstract}
A physics-based reduced-order model unifies curved-meniscus inertia, viscous losses and nonsmooth contact-line stick--slip in a passive port-Hamiltonian formulation for lateral sloshing in cylindrical tanks. Dirichlet--Neumann and Jacobi operators describe inertia and restoring forces, while a convex potential combines bulk, Stokes-layer and contact-line dissipation. Analytical propagation between stick--slip events supports faster-than-real-time simulation of coupled spacecraft translation and sloshing. The viscous closure agrees within $5\,\%$ with a reference Stokes calculation, and the model reproduces the Henderson--Miles modal measurements and the Cocciaro hysteretic decay. Across 42 drop-tower and ground measurements, curvature correction reduces the root-mean-square frequency error from $24$ to $4\,\%$. For flight and ground tests from the Middeck 0-Gravity Dynamics Experiment, frequency errors range from $0.9$ to $7.1\,\%$, while ideal contact lines underestimate the damping ratios from the MODE fits by $38$--$87\,\%$. One hysteresis width, calibrated on the lowest orbital water excitation, predicts higher orbital excitations and the ground test with normalised force errors of $11$--$35\,\%$, against $149$--$208\,\%$ for an ideal pinned line. It predicts $8.5\,\%$ resonance softening against $7.3\,\%$ measured, which the tested one-parameter damping laws miss. Uncertain wetting and remaining response discrepancies motivate damping ranges in spacecraft control analyses.
\end{abstract}


\hypersetup{pdfauthor={Gianni Cassoni, Fabrizio Gennari, Jacopo Serafini, Pierangelo Masarati}}

\section{Introduction}
\label{sec:intro}

Liquid propellant can represent a large fraction of spacecraft mass, and its motion couples to attitude control, docking and in-orbit refuelling \citep{abramson1966,simonini2024}. Guidance, navigation and control (GNC) design relies on fast, energy-consistent equivalent pendulum and mass--spring models, whose parameters follow from flat-interface theory when the Bond number $\Bo=\rho gR^2/\sigma$ is large, with $\rho$ the liquid density, $\sigma$ the surface tension, $g$ the axial acceleration and $R$ the tank radius. In orbit the Bond number of a $0.5\,$m water tank falls from about $10^4$ to below one. The liquid then rests beneath a curved meniscus selected by the static contact angle $\theta_c$ \citep{kopachevskii1973}, and the parameters require curved-interface modelling or calibration \citep{utsumi2004,utsumi2014,elke2024}. Flight and drop-tower campaigns, from the NASA drop-tower tests \citep{salzman1969} and the Middeck 0-Gravity Dynamics Experiment (MODE) \citep{mode1993,vanschoor1995} to Sloshsat-FLEVO, SPHERES-Slosh, FLUIDICS and ferrofluid experiments \citep{gerrits2003,dalmon2019,romerocalvo2021,romerocalvo2022}, highlight uncertainties in contact-line modelling \citep{storey2017,silveri2026}. Meniscus-dependent sloshing also couples to spacecraft pitching and flexible appendages \citep{chiba2021}. Interface-resolving computational fluid dynamics (CFD), when validated for the relevant flow regime, can simulate spacecraft manoeuvres \citep{dalmon2019}, although its computational cost can limit repeated GNC analyses and real-time applications.

Curved-interface spectra and edge constraints have established low-gravity formulations \citep{shankar2003,shankar2007,utsumi2008,benjaminscott1979,prosperetti2012}. In Zakharov--Craig--Sulem variables \citep{zakharov1968,craig1993}, curvature appears in two explicit operators, the Dirichlet--Neumann (DN) map of the curved domain and the Jacobi operator of the meniscus. The conservative formulation of \citet{cassoni2026} recovers the curved-meniscus benchmark trends of \citet{kopachevskii1973} and separates the kinetic and capillary contributions to wetting/non-wetting asymmetry. Damping and forced response add boundary layers \citep{miles1967}, corner and capillary layers \citep{miles1991,hocking1987}, and contact-line friction and hysteresis \citep{cocciaro1993,viola2018}. \citet{utsumi2013,utsumi2017} added the work of the wall boundary layer and of a contact-line friction force to the low-gravity modal equation and found that friction damping grows markedly as the Bond number decreases. Port-Hamiltonian shallow-water models couple sloshing and flexible structures \citep{cardoso2017}. Viscous slip-length models with a hysteretic contact law \citep{violagallaire2018prf} and successive modal projections for stick--slip relaxation \citep{bongarzone2024stick} cover parts of this physics.

This work extends the conservative curved-meniscus formulation of \citet{cassoni2026} to dissipative lateral sloshing in low-Bond cylindrical tanks, incorporating bulk and Stokes-layer losses together with nonsmooth contact-line hysteresis. The same curved-domain harmonic fields determine inertia and viscous dissipation, while a convex contact-line potential introduces linear resistance and stick--slip dynamics within a port-Hamiltonian reduced-order model. The inclusion is passive on the admissible discrete space and propagated analytically between stick--slip events. Unlike the friction-work model of \citet{utsumi2017}, it resolves stick phases and their effects on frequency, force phase and damping. One width calibrated on the lowest orbital MODE excitation predicts two higher excitations and the ground response, including softening missed by the tested linear and quadratic damping laws. A design chart supplies frequencies and slosh masses, with frequencies checked against independent drop-tower data. Pole ranges quantify contact-line sensitivity across gravity levels. 

\section{Model and methods}
\label{sec:formulation}

\subsection{Energy of the curved free surface}

\begin{figure}[!tb]
\centering
\includegraphics[width=0.85\textwidth]{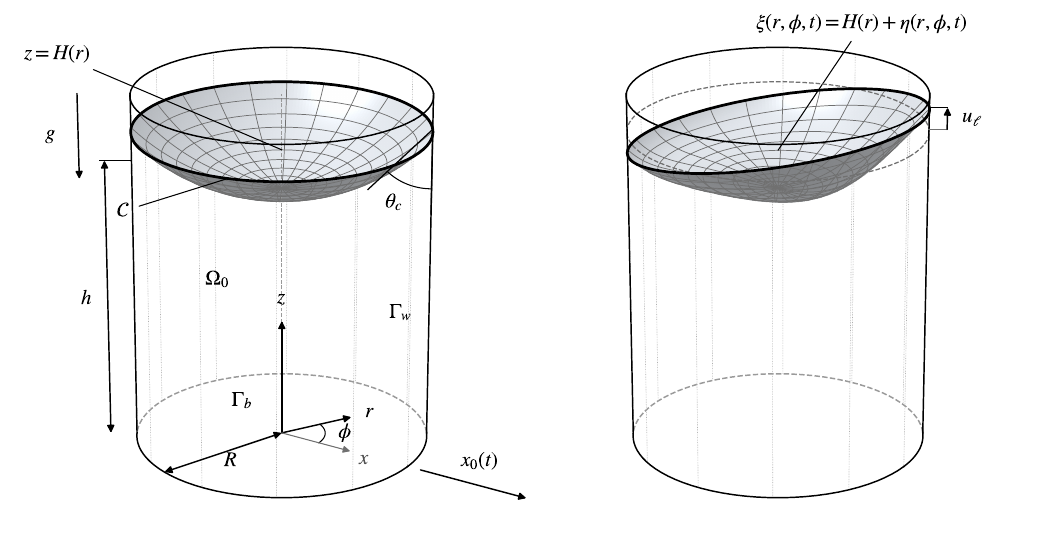}
\caption{Geometry and notation for the MODE water tank in orbit ($R=15.5\,$mm, $h=31\,$mm, $\theta_c=45^\circ$, $\Bo=0$). Left: equilibrium meniscus and tank displacement. Right: perturbed interface $\xi=H+\eta$ (first two mobile-line modes, exaggerated), with the equilibrium contact line dashed.}
\label{fig:geometry}
\end{figure}

A rigid circular cylinder of radius $R$ with a flat bottom contains liquid of density $\rho$, kinematic viscosity $\nu$ and surface tension $\sigma$, with mean depth $h$ (Fig.~\ref{fig:geometry}). Cylindrical coordinates $(r,\phi,z)$ have $z$ upward from the bottom and $\phi$ measured from the direction $x$ in which the tank is translated by $x_0(t)$, and the axial acceleration $g\ge0$ points towards the bottom. The liquid occupies the domain $\Omega_0$, bounded by the wetted wall $\Gamma_w$ and the bottom $\Gamma_b$. The flow is irrotational and the free surface is a graph $z=\xi(r,\phi,t)$ over the disc $\mathcal{S}=\{r\le R\}$. In the variables of \citet{zakharov1968} and \citet{craig1993}, the state is the elevation $\xi$ and the surface potential $\psi$, and the bulk enters only through the Dirichlet--Neumann (DN) operator $\Gop[\xi]$. It returns the normal derivative of the velocity potential $\Phi$ at the surface scaled by the area factor $\sqrt{1+|\nabla\xi|^2}$, that is, the flux per unit projected area, with $\Phi$ the harmonic extension of $\psi$ into the liquid with no flux through the wall and bottom. The energy per unit density is
\begin{equation}
\Ham(\xi,\psi)=\underbrace{\tfrac12\!\int_{\mathcal{S}}\psi\,\Gop[\xi]\psi\,\dd A}_{\text{kinetic}}
+\underbrace{\tfrac g2\!\int_{\mathcal{S}}\xi^2\,\dd A}_{\text{gravitational}}
+\underbrace{\kappa\!\int_{\mathcal{S}}\bigl(\sqrt{1+|\nabla_s\xi|^2}-1\bigr)\dd A}_{\text{capillary}}
\,\underbrace{-\,\kappa\cos\theta_c\!\oint_{\mathcal{C}}\xi\,\dd\ell}_{\text{wetting}},
\label{eq:hamiltonian}
\end{equation}
with $\kappa=\sigma/\rho$, $\nabla_s$ the gradient on $\mathcal{S}$, $\theta_c$ the static contact angle and $\mathcal{C}$ the contact line $r=R$, where the free surface meets the wall. The wetting energy is $-\kappa\cos\theta_c$ times the wetted wall area, which on the vertical wall of the cylinder equals $\oint_{\mathcal{C}}\xi\,\dd\ell$. The motion is Hamiltonian, $\pp_t\xi=\delta_\psi\Ham$, $\pp_t\psi=-\delta_\xi\Ham$ \citep{zakharov1968,craig1993}.

At rest $\psi=0$ and the surface is $z=H(r)$. Writing $\xi=H+\eta$ and expanding \eqref{eq:hamiltonian} in powers of the small perturbation $(\eta,\psi)$,
\begin{equation*}
\Ham(H+\eta,\psi)=\Ham_0+\Ham_1+\Ham_2+O\bigl(|(\eta,\psi)|^3\bigr),
\end{equation*}
where $\Ham_0$ is the energy of the rest state, a constant that does not affect the motion, and $\Ham_k$ is of order $k$. The rest state is an equilibrium when it is a stationary point at fixed volume, that is, when the first-order energy vanishes. After integration by parts,
\begin{equation*}
\Ham_1=\int_{\mathcal{S}}\Bigl[gH-\frac{\kappa}{r}\Bigl(\frac{rH'}{\Lambda}\Bigr)'-p_0\Bigr]\eta\,\dd A
+\kappa\oint_{\mathcal{C}}\Bigl(\frac{H'}{\Lambda}-\cos\theta_c\Bigr)\eta\,\dd\ell,\qquad \Lambda=\sqrt{1+H'^2},
\end{equation*}
where $p_0$ is the Lagrange multiplier that enforces the volume. Requiring $\Ham_1=0$ for every $\eta$ gives the Young--Laplace equation in the interior and Young's law at the wall,
\begin{equation}
\frac{\kappa}{r}\frac{\dd}{\dd r}\Bigl(\frac{rH'}{\Lambda}\Bigr)=gH-p_0,\qquad H'(0)=0,\qquad H'(R)=\cot\theta_c,
\label{eq:yl}
\end{equation}
where $H'(0)=0$ is the regularity condition on the axis and the wall condition, $H'(R)/\Lambda=\cos\theta_c$, balances the capillary pull of the free surface against the wetting force. Equation~\eqref{eq:yl} is solved by slope-angle shooting \citep{concus1968}, or volume-constrained collocation for near-wetting benchmarks. On the vertical wall of a cylinder the wetted area is linear in $\xi$, so the wetting term fixes the contact angle but stores no energy in small motions. On walls whose area per unit height varies, as in conical or domed tanks, it would add a contact-line stiffness.

With $\Ham_1=0$ at equilibrium, and no wetting contribution on the vertical wall, the leading term of the expansion is the second-order energy $\Ham_2$. The kinetic term is already quadratic in $\psi$, so $\Gop$ is evaluated on the equilibrium surface, and expanding the capillary term to second order gives
\begin{equation}
\Ham_2(\eta,\psi)=\frac12\int_{\mathcal{S}}\psi\,\Gop[H]\psi\,\dd A
+\frac12\int_{\mathcal{S}}\Bigl[g\,\eta^2+\kappa\Bigl(\frac{\eta_r^2}{\Lambda^3}+\frac{\eta_\phi^2}{r^2\Lambda}\Bigr)\Bigr]\dd A .
\label{eq:H2}
\end{equation}
Curvature enters in two places: in the DN operator $\Gop[H]$ of the curved domain, which sets the inertia, and in the metric factor $\Lambda$, which weights the capillary stiffness. The gravity and capillary terms of \eqref{eq:H2} define the Jacobi operator $\Jop[H]$ of the meniscus. The two classical contact-line conditions are the two boundary conditions admitted by \eqref{eq:H2}. On the vertical cylindrical wall, a mobile line preserving the equilibrium angle satisfies the linearised natural condition $\eta_r(R)=0$. A line pinned at its equilibrium height satisfies $\eta(R)=0$. Prescribed contact-angle conditions need not be compatible with smooth inviscid motion \citep{shankarkidambi2005}. Here, fixed-angle motion is a variational closure for the ideal mobile-line limit \citep{cassoni2026}. The pinned and hysteretic conditions allow the angle to vary.

\subsection{Dissipation and contact-line hysteresis}

All losses are collected in one convex dissipation potential $\Dis$ of the velocity potential $\Phi$ and of the contact-line speed $u_\ell=\pp_t\eta(R,\phi,t)$:
\begin{equation}
\Dis=\underbrace{\nu\int_{\Omega_0}\nabla\nabla\Phi:\nabla\nabla\Phi\,\dd V}_{\text{bulk}}
+\underbrace{\frac{\beta(\omega)}{2}\int_{\Gamma_w\cup\Gamma_b}|\nabla_{\!\tau}\Phi|^2\,\dd S}_{\text{Stokes layers}}
+\frac1\rho\oint_{\mathcal{C}}\Bigl(\underbrace{\tfrac12\hat c\,u_\ell^2}_{\mathclap{\substack{\text{linear}\\\text{resistance}}}}+\underbrace{\hat f\,|u_\ell|}_{\mathclap{\text{hysteresis}}}\Bigr)\dd\ell.
\label{eq:dissipation}
\end{equation}
The bulk term represents irrotational viscous loss in $\Omega_0$, $\nabla_{\!\tau}$ in the second term is the gradient tangential to the wall and bottom, $\omega$ is the angular frequency of the motion and $\beta(\omega)=\sqrt{\nu\omega/2}$ is the Stokes-layer coefficient. For harmonic motion, thin Stokes layers of thickness $\sqrt{2\nu/\omega}$ correct the wall and bottom slip and have equal dissipative and reactive impedance components \citep{miles1967,henderson1998}. At rest their loss vanishes with velocity. The thin-layer approximation does not extend to $\omega=0$.

The contact-line term combines a linear resistance $\hat c=\rho\nu\,\alpha\sin\theta_c$ with friction coefficient $\alpha$. It corresponds to the law $\theta-\theta_c\simeq\alpha\,\Ca_\ell$ of \citet{viola2018}, with $\theta$ the dynamic contact angle and the contact-line capillary number $\Ca_\ell=\rho\nu u_\ell/\sigma$, which is the small-$\Ca_\ell$ linearisation of the Cox--Voinov relation \citep{cox1986,snoeijer2013}. At the line speeds of the present tanks, $\Ca_\ell\sim10^{-4}$, contact-line motion lies near the crossover between the hydrodynamic and the thermally activated regimes \citep{snoeijer2013}, so a constant $\alpha$ is an approximation. This mobility has been measured independently on vibrated sessile drops and then used without fitting to predict drop coalescence \citep{xiasteen2018,ludwicki2022}, and for water these measurements give $\alpha\approx200$--$600$.

The model represents hysteresis by a symmetric Young-force threshold of magnitude $\hat f=\tfrac{\sigma}{2}(\cos\theta_R-\cos\theta_A)$, where $\theta_A$ and $\theta_R$ are the advancing and receding contact angles. This approximation retains the width of the Young-force interval and centres it at zero. It follows the dry-friction picture of hysteresis as pinning on surface defects \citep{snoeijer2013} and the Coulomb analogy for sloshing contact lines \citep{miles1967,viola2018}. Together the two parts form the Davis--Hocking law with separate advancing and receding angles, the contact law also used in the viscous weakly nonlinear sloshing analysis of \citet{violagallaire2018prf}. Whether they act is set by an inclusion: the unbalanced Young force per unit length $f_\ell$ must lie in the subdifferential of the line potential,
\begin{equation}
f_\ell\in\hat c\,u_\ell+\hat f\,\Sgn(u_\ell)=
\begin{cases}
[-\hat f,\hat f], & u_\ell=0\ \text{(stick)},\\
\hat c\,u_\ell+\hat f\sgn u_\ell, & u_\ell\neq0\ \text{(slip)},
\end{cases}
\label{eq:line_inclusion}
\end{equation}
where $\Sgn$ is the set-valued sign, equal to $\sgn$ for nonzero arguments and with $\Sgn(0)=[-1,1]$. While $|f_\ell|<\hat f$ the line sticks and neither term dissipates. Once $|f_\ell|$ reaches $\hat f$ it can slip, and both terms then act together. The ideal classes are the limits $\hat c=\hat f=0$ (mobile, or free, line: $f_\ell=0$) and $\hat f\to\infty$ (pinned, permanent stick), and the set-valued stick branch makes the dynamics nonsmooth. Every term of \eqref{eq:dissipation} removes energy, so $\Ham_2$ can grow only through the power supplied by the tank motion. Free-surface viscous stresses, corner flows, Marangoni stresses and thin films are omitted from the baseline closure.

\subsection{Reduced-order model}

Projecting onto $N$ surface coordinates in the lateral sector of azimuthal wavenumber $m=1$, $\eta=\cos\phi\sum_jS_j(r)q_j(t)$, reduces \eqref{eq:H2} and \eqref{eq:dissipation}, per unit $\rho\pi$, to
\begin{equation}
\Ham=\tfrac12\qd\tp\T\qd+\tfrac12\q\tp\K\q,\qquad
\Dis=\tfrac12\qd\tp\D\qd+\tfrac12c_\ell(\e\tp\qd)^2+F_h|\e\tp\qd| .
\label{eq:discrete_functionals}
\end{equation}
The inertia $\T=\M\G^{-1}\M$, with $\M$ the surface mass matrix and $\G$ the DN matrix, carries the curved-domain kinetic coupling. The stiffness is $\K=g\M+\kappa\Lc$, with $\Lc$ the capillary form. The viscous damping $\D=\A\tp\C_{\rm s}\A+\beta\Pim$, with $\A=\G^{-1}\M$, uses the bulk Gram $\C_{\rm s}$ of the harmonic lifts of the surface basis and the boundary-layer Gram $\Pim$ of the same fields. The inertia and viscous forms are computed through bulk harmonic lifts, whereas the stiffness is computed on the free surface. The vector $\e$ returns the contact-line height $\e\tp\q$. Integrating the local contact law around the ring gives
\begin{equation}
c_\ell=R\,\alpha\,\nu\sin\theta_c,\qquad F_h=\frac{2R\kappa}{\pi}\bigl(\cos\theta_R-\cos\theta_A\bigr).
\label{eq:contact_coeffs}
\end{equation}
These coefficients follow from restricting the line potential of \eqref{eq:dissipation} to the motion $u_\ell=u\cos\phi$ of the sector $m=1$, through $\oint\cos^2\phi\,\dd\ell=\pi R$ and $\oint|\cos\phi|\,\dd\ell=4R$. Convexity and dissipation are preserved under this restriction. The reduction enforces a single stick--slip state for the whole ring, whose velocity varies as $\cos\phi$, so independent depinning at different azimuths is excluded (\S\ref{sec:hysteresis}). The Lagrange--d'Alembert principle with the convex potential \citep{moreau1988} then yields the governing system
\begin{equation}
\begin{aligned}
&\T\qdd+\D\qd+\K\q+\e\,\lambda=-\bb\,\ddot x_0,\\
&\lambda-c_\ell u\in F_h\,\Sgn(u),\qquad u=\e\tp\qd,\\
&F_x=m_f\,\ddot x_0+\rho\pi\,\bb\tp\qdd,
\end{aligned}
\label{eq:dvi_system}
\end{equation}
where $\lambda$ is the contact-line reaction, $u$ the reduced line velocity, $b_j=\int_0^RS_jr^2\dd r$, $m_f$ the liquid mass and $F_x$ the force that accelerates the liquid with the tank (the tank reaction used in phase plots is $-F_x$). Harmonic responses use the frequency-dependent Stokes terms directly. Modal and time-domain calculations freeze them at a reference frequency $\omega_*$, normally the tracked modal frequency, and there $\T$ and $\D$ stand throughout for $\T_*=\T+\beta(\omega_*)\Pim/\omega_*$, which includes the reactive Stokes-layer term, and $\D_*=\A\tp\C_{\rm s}\A+\beta(\omega_*)\Pim$. The frozen terms reproduce the harmonic impedance at $\omega_*$ and preserve passivity in time. With the momenta $\p=\T\qd$, the state $\x=(\q,\p)$ and the generalised force $\f=-\bb\,\ddot x_0$, \eqref{eq:dvi_system} is a port-Hamiltonian system with a nonsmooth resistive port,
\begin{equation}
\begin{gathered}
\dot\x=(\Jm-\Rm)\nabla\Ham-\Bm\e\lambda+\Bm\f,\quad
\Jm=\begin{bmatrix}\mat0&\I\\-\I&\mat0\end{bmatrix},\
\Rm=\begin{bmatrix}\mat0&\mat0\\\mat0&\D\end{bmatrix},\
\Bm=\begin{bmatrix}\mat0\\\I\end{bmatrix},\\
\dot\Ham=-\qd\tp\D\qd-u\lambda+\qd\tp\f\le\qd\tp\f .
\end{gathered}
\label{eq:ph}
\end{equation}
Since $\D\succeq0$ and $u\lambda\ge0$, the frozen model is passive for every admissible truncation with positive-definite inertia and non-negative stored energy. In harmonic responses $\D(\omega)$ is positive semidefinite at each real frequency, so the dissipation averaged over a cycle is non-negative. In the spacecraft demonstration, halving $\omega_*$ changes peak force by about $-1\,\%$ and increases settling times by $30$--$43\,\%$. Doubling it changes peak force by $+1\,\%$ and reduces settling times by $19$--$25\,\%$. Peak force is relatively insensitive, but transient decay is most reliable when one frequency dominates. Broadband transients require a causal impedance realisation.

The two ideal limits of \eqref{eq:dvi_system} are the mobile line ($F_h=c_\ell=0$, natural condition $\eta_r(R)=0$) and the pinned line ($F_h\to\infty$, essential condition $\eta(R)=0$). For the MODE water tank in orbit, a pinned contact ring stays fixed while the interface sloshes inside it, at an inviscid $f_1=3.14\,$Hz, whereas a mobile ring tilts as a whole at $1.50\,$Hz. Neither limit dissipates at the contact line, because a free line meets no resistance and a pinned line does not move. Line dissipation exists only in the resistive regime between them, which is central to the MODE damping in \S\ref{sec:mobility}.

Free modes $\q=\vv e^{st}$, with complex pole $s$, are reported through $f_n=|s|/2\pi$, the decay rate $\gamma=-\operatorname{Re}s$ and $\zeta=\gamma/|s|$. For harmonic tank motion with an ideal or linear-resistive contact line, \eqref{eq:dvi_system} gives the force transfer function, in terms of the complex amplitudes $\hat F_x$ and $\hat x_0$,
\begin{equation}
\frac{\hat F_x}{-m_f\omega^2\hat x_0}=1+\frac{\rho\pi\omega^2}{m_f}\,\bb\tp\bigl(\K-\omega^2\T(\omega)+\mathrm{i}\omega\D(\omega)\bigr)^{-1}\bb
\approx1+\sum_n\mu_n\frac{\omega^2}{\omega_n^2-\omega^2+2\mathrm{i}\zeta_n\omega_n\omega},
\label{eq:mass_fraction}
\end{equation}
with $\T$-orthonormal modes $\vv_n$, natural angular frequencies $\omega_n=2\pi f_n$, damping ratios $\zeta_n$ and slosh-mass fractions $\mu_n=\rho\pi(\vv_n\tp\bb)^2/m_f$. The one-mode form corresponds to the two-complex-pole fit of \citet{vanschoor1995}, whose participating mass fraction is comparable with $\mu_1$. For a pinned line, the transfer function is evaluated by projecting the matrices and forcing vector onto the subspace $\e\tp\q=0$. A resistive line adds $c_\ell\e\e\tp$ to $\D$. For hysteresis, harmonic balance uses $c_{\rm eq}$ in place of $c_\ell$, giving an approximate, amplitude-dependent response.

The potential is expanded in harmonic lifts $\Phi_k=J_1(k_kr)\cosh(k_kz)\cos\phi$ with $J_1'(k_kR)=0$. These lifts satisfy the rigid-wall conditions exactly. Laboratory benchmarks use the corresponding $J_m$ sector, conserving volume for $m=0$. Bessel mobile surfaces, wall-vanishing Jacobi polynomials and an edge function $r/R$ span the mobile, pinned and resistive classes. Tensor Gauss--Legendre quadrature evaluates the bulk and boundary Grams. QR factorisation forms $\T$ without explicit inversion. Across $108$ surface, bulk and quadrature refinement comparisons, frequencies changed by at most $0.067\,\%$ and damping ratios by $0.19\,\%$. Each linear stick or slip regime is propagated analytically, with events located by Brent's method without regularisation. Accumulated work closes \eqref{eq:ph} to round-off.

For total spacecraft mass $M$, displacement $x$ and thrust $F_{\rm thr}$, momentum conservation gives $M\ddot x+\rho\pi\bb\tp\qdd=F_{\rm thr}$. Eliminating $\ddot x$ gives the reduced inertia $\T-\rho\pi\bb\bb\tp/M$, which is positive definite when $M>\rho\pi\bb\tp\T^{-1}\bb$. The 24-state MODE demonstration runs five times faster than real time (propagation only), with an energy residual of about $0.1\,\%$ of peak energy. Underestimating damping is not necessarily conservative for attitude control, because internal liquid dissipation can drive nutation growth during minor-axis rotation \citep{hubert2008}. This rotational effect is outside the translational demonstration.

\subsection{Data and error measures}
\label{sec:data}

Tabulated values are used as published, and force and phase responses are digitised from the published figures by manual picks, whose pixel uncertainty is reported as the extraction error. Force errors are root-mean-square deviations normalised by mean measured gain (NRMSE). Phase errors use wrapped differences. 

The viscous closure is checked against the linearised Navier--Stokes (LNS) decay rates of \citet{bongarzone2022slip} and the six clean-water pinned modes of \citet{henderson1994}, as tabulated by \citet{martel1998}, through the damping parameter $\Delta=2R\gamma/\sqrt{\pi\nu f_0^{\rm HM}}$, with $f_0^{\rm HM}$ their inviscid reference frequency. The hysteresis inclusion is tested on the decay curve of \citet{cocciaro1993} as reproduced by \citet{viola2018}, with their contact law unchanged.

The design chart is checked against $42$ deep-liquid ($h/R>2$) frequencies of \citet{salzman1969}, measured in the NASA Lewis five-second drop tower and in normal gravity for seven fully wetting liquids up to $\Bo=835$, through the frequency parameter $\Omega^2=\omega_1^2\rho R^3/\sigma$, with $\omega_1$ the first-mode angular frequency.

The flight data come from MODE on STS-48 \citep{mode1993,vanschoor1995}, whose cylinders, of radius $15.5\,$mm and filled to about twice the radius, held distilled water or silicone oil, with fluid properties and fills from Tables~2 and 3 of \citet{vanschoor1995}. Tables~5 (oil) and 6 (water) provide first-mode frequency, damping and slosh mass. Force responses at $x_0/d=0.32$, $1.02$ and $3.22\,\%$, where $d=2R$ is the tank diameter, are digitised from Figs.~9, 10, 15 and 16 of \citet{vanschoor1995}. Table~\ref{tab:hysteresis} uses the journal phase data, and Fig.~\ref{fig:mode_sensitivity} uses the corresponding phase series from the NASA MODE report \citep{mode1993}. Contact angles, hysteresis, residual acceleration and dwell schedules were not reported, so the baseline assumes static angles of $45^\circ$ for water and $30^\circ$ for oil, with ground fills equal to the space fills. The contact-line classes follow the experimenters, who attributed the water behaviour to contact-angle hysteresis \citep[p.~530]{vanschoor1995}:
\begin{quote}
\emph{``The first-order effect of contact-angle hysteresis is to increase the slosh frequency, and since contact-angle hysteresis is not a function of the ambient acceleration, the frequency change is smaller when gravity is removed. It can be inferred that in space the contact-angle hysteresis dominates the restoring force, keeping the frequency high, the damping ratio low, and the nonlinearity present.''}
\end{quote}
Silicone oil, which wets almost any solid and shows little hysteresis, was found to be essentially linear and highly damped in space. The baseline (pinned water, mobile oil) follows this interpretation, and \S\ref{sec:bands} tests it independently.


\FloatBarrier

\section{Results}
\label{sec:results}

\FloatBarrier

\subsection{Viscous closure and laboratory benchmarks}
\label{sec:lns}

The viscous closure is first checked on the configuration of \citet{bongarzone2022slip}, run without modification. The ROM lies $2$--$12\,\%$ below their linearised Navier--Stokes (LNS) decay rates, with the larger deficit at the smallest radii. The effect of meniscus curvature was checked separately using a mesh-converged Taylor--Hood finite-element solution of the linear Stokes equations for the MODE water geometry. The calculation retains the lateral $m=1$ sector, a pinned contact line, no slip at the wall and bottom, and a clean stress-free interface. A Henderson--Miles control case gives errors of $0.35\,\%$ in frequency and $2.8\,\%$ in damping parameter $\Delta$. For flat and curved menisci under orbital and ground gravity, the ROM agrees with the Stokes frequencies and underpredicts decay rates by at most $5\,\%$. Thus, the viscous effects omitted by the ROM, including the contact-line corner flow, produce only a small net correction in these clean-surface cases.

\label{sec:cocciaro}
\begin{figure}[!tb]
\centering
\includegraphics[width=0.78\textwidth]{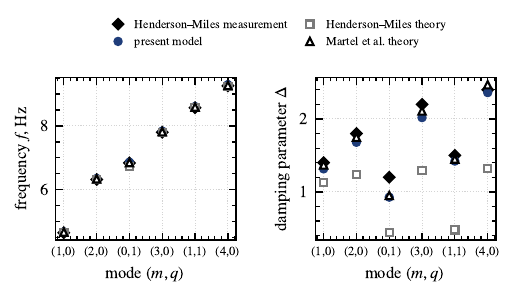}
\caption{Henderson--Miles brimful cylinder (clean water, flat interface, pinned line, $R=27.66\,$mm, $h=38\,$mm): frequency (left) and damping parameter $\Delta$ (right) of the six modes $(m,q)$, with azimuthal index $m$ and radial index $q$. Measurements from Table~3 of \citet{martel1998} (diamonds), present model (circles), theory of \citet{henderson1994} (squares) and of \citet{martel1998} (triangles). No parameter is fitted.}
\label{fig:henderson}
\end{figure}

For the six Henderson--Miles modes (Fig.~\ref{fig:henderson}), the frequency error is at most $0.55\,\%$ and the damping parameter $\Delta$ has a mean error of $8.4\,\%$ (maximum $22.7\,\%$), comparable with the corrected theory of \citet{martel1998} and consistent with the deficit found against LNS.

\begin{figure}[!tb]
\centering
\includegraphics[width=0.78\textwidth]{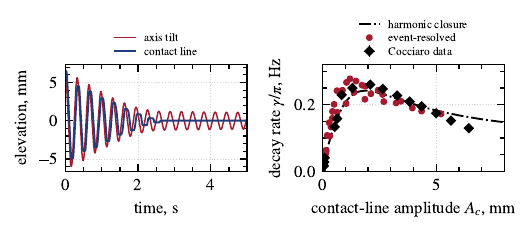}
\caption{Cocciaro configuration ($R=50.25\,$mm, $h/R=2.58$, $\theta_c=62^\circ$, $\theta_A-\theta_R=43^\circ$, $\alpha=150$, from \citet{viola2018}). Left: event-resolved free decay. The contact-line elevation (navy) sticks in plateaus and stops near $2.6\,$s, while the axis tilt (crimson) keeps oscillating. Right: decay rate against contact-line amplitude, measured by \citet{cocciaro1993} (diamonds) and computed with the event-resolved inclusion (circles) and the harmonic closure (dash-dotted). No parameter is fitted.}
\label{fig:cocciaro}
\end{figure}

Against the viscous curved-meniscus calculations of \citet{kidambi2009wave}, digitised from Fig.~12 of \citet{bongarzone2022pinned}, with depth taken from the bottom to the contact ring, the pinned ROM has mean errors below $0.3\,\%$ in frequency and $4$--$6\,\%$ in decay rate, without fitted coefficients.

For the hysteretic decay of \citet{cocciaro1993}, the contact law is used as reported and the decay is evaluated from the tilt at the axis, as in the experiment. The line slips during large oscillations, sticks for progressively longer intervals and finally stops, while the liquid keeps oscillating in pinned modes (Fig.~\ref{fig:cocciaro}). The decay matches the measurement with an NRMSE of $8.8\,\%$ above $2\,$mm, comparable with the $10\,\%$ extraction uncertainty. Below $2\,$mm the inclusion also follows the measured fall of the decay rate, which \citet{viola2018} attributed to pinning and could not capture with their free-edge amplitude equation. The harmonic closure $c_{\rm eq}=4F_h/(\pi\omega A_c)$, the first harmonic of the dry-friction term at contact-line amplitude $A_c$, gives a similar error above $2\,$mm. The collective-ring approximation excludes independent local depinning. Under a cosine load, local slip begins at $\pi/4$ of the collective threshold. Applying this factor raises the Cocciaro decay error from $9$ to $16\,\%$ and the calibrated MODE width from $27$ to $34^\circ$.

\begin{table}[!tb]
\centering
\caption{MODE first mode: model with ideal contact lines (pinned water, mobile oil) against first-mode fits to the measurements \citep{vanschoor1995}. $\mu_1$ is the slosh-mass fraction. The orbital oil damping is only weakly constrained by its broad response (\S\ref{sec:hysteresis}).}
\label{tab:validation}
\begin{adjustbox}{max width=\textwidth}\begin{tabular}{@{}llrrrrrrrr@{}}
\toprule
Fluid & Env. & $f_n$ model (Hz) & meas. & error & $\zeta$ model & meas. & error & $\mu_1$ model & meas.\\
\midrule
Water & orbit & 3.113 & 3.31 & $-5.9\,\%$ & 0.0108 & 0.044 & $-75.6\,\%$ & 0.179 & 0.17\\
Oil   & orbit & 0.686 & 0.68 & $+0.9\,\%$ & 0.1005 & 0.363 & $-72.3\,\%$ & 0.210 & 0.24\\
Water & ground & 6.744 & 6.30 & $+7.1\,\%$ & 0.0060 & 0.047 & $-87.2\,\%$ & 0.197 & 0.26\\
Oil   & ground & 5.325 & 5.24 & $+1.6\,\%$ & 0.0249 & 0.040 & $-37.6\,\%$ & 0.251 & 0.24\\
\bottomrule
\end{tabular}\end{adjustbox}
\end{table}

\subsection{Curvature corrections for equivalent mechanical models}
\label{sec:design}

\begin{figure}[!tb]
\centering
\includegraphics[width=0.72\textwidth]{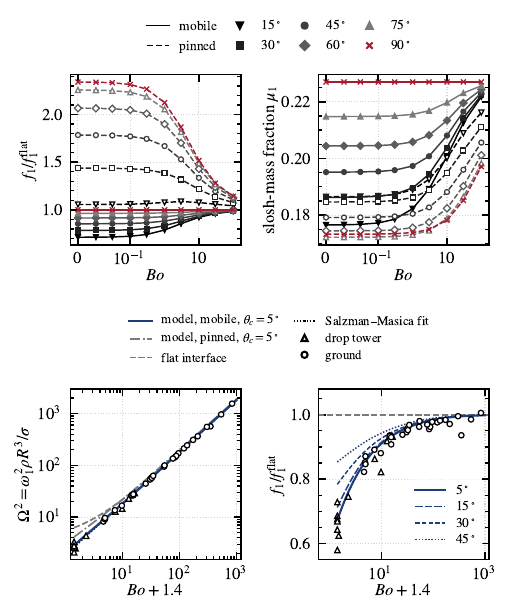}
\caption{Design chart for $h/R=2$, mobile and pinned lines and several static angles: Upper panels: first natural frequency relative to the flat-interface value (left) and slosh-mass fraction (right). Lower panels: the $42$ frequencies of \citet{salzman1969} (triangles: drop tower, circles: ground) against the mobile-line model. No parameter is fitted.}
\label{fig:design}
\end{figure}

The conservative formulation provides the first natural frequency $f_1$ and participating slosh-mass fraction $\mu_1$, which determine the stiffness and moving mass of a first-mode mechanical analogue \citep{utsumi2004,utsumi2008}. The upper panels of Fig.~\ref{fig:design} show how these quantities vary with Bond number and static contact angle at $h/R=2$, with frequency normalised by $f_1^{\rm flat}$ at the same Bond number. At $90^\circ$ the mobile line reproduces the flat-interface values exactly. At low Bond number a strongly wetting mobile meniscus lowers the frequency by almost a third and the slosh mass by about a fifth, while pinning more than doubles the frequency and lowers the slosh mass further. Both corrections fade once the Bond number exceeds a few tens. For propellant tanks in orbit, flat-interface models can therefore misplace the slosh frequency by tens of per cent and overestimate the slosh mass by up to a quarter.

The lower panels of Fig.~\ref{fig:design} compare the mobile-line predictions with $42$ drop-tower and ground measurements from \citet{salzman1969}. The equilibrium and ROM are computed at each measured Bond number using a fixed static angle of $5^\circ$. The overall RMS frequency error is $4\,\%$, compared with $24\,\%$ for a flat interface and approximately $4\,\%$ for the report's empirical correlation. Curvature correction matters most at low Bond number. For $\Bo<0.01$, $0.01$--$10$, $10$--$100$ and $\Bo>100$, the ROM errors are respectively $8$, $5$, $1.3$ and $2.4\,\%$, compared with $53$, $27$, $6$ and $3\,\%$ for a flat interface.

Pinning raises the predicted frequency by about $20\,\%$, worsening agreement with these measurements. The data therefore support the mobile-line approximation for these fully wetting liquids, which is also adopted below for MODE oil. The reported logarithmic decrements are excluded from the damping comparison because the experiments involve a residual wall film \citep{salzman1969}, whose dynamics are absent from the ROM.

\FloatBarrier

\subsection{The MODE flight experiment}
\label{sec:mode}

The MODE comparison examines linear contact-line resistance and hysteresis. Linear resistance is calibrated separately to each measured pole, while one hysteresis width is calibrated on the lowest orbital force curve and used to predict the remaining responses. These held-out cases share hardware and processing uncertainties, so differences in their preferred widths indicate limits of predictive transfer rather than independent measurements of hysteresis.

\FloatBarrier

\subsubsection{Frequencies, slosh mass and viscous damping}
\label{sec:bands}

\begin{figure}[!tb]
\centering
\includegraphics[width=0.56\textwidth]{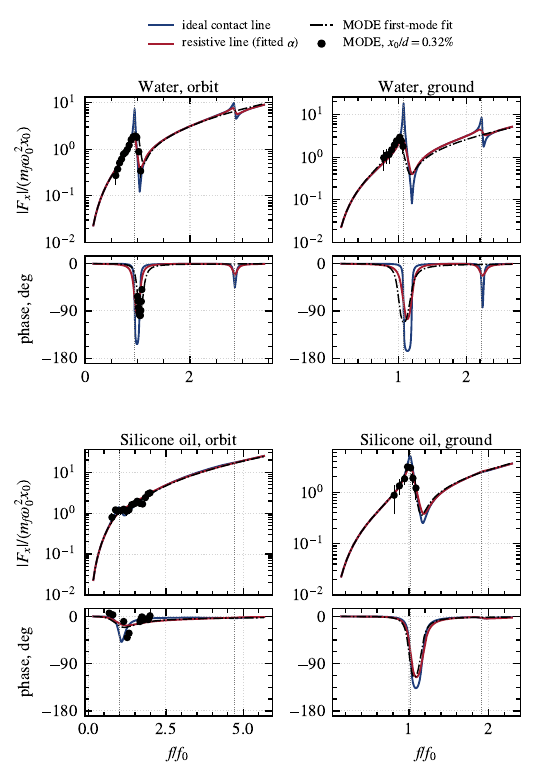}
\caption{MODE force spectra up to the second radial mode, magnitude above phase for each case, against the frequency normalised by the measured first natural frequency $f_0$. Dotted lines mark the model natural frequencies. The ground phase was not digitised.}
\label{fig:spectra}
\end{figure}

Table~\ref{tab:validation} compares the first mode with MODE. Frequencies agree within $0.9$--$7.1\,\%$. The slosh-mass fraction, which tests the mode shape independently of frequency, agrees within about $5\,\%$ for orbital water and ground oil, and is lower for orbital oil and for ground water, whose fill is assumed. The damping ratio, by contrast, is $38$--$87\,\%$ too low. The force spectra of Fig.~\ref{fig:spectra} show the consequence: the model places the resonances correctly, but with ideal contact lines the water resonances are far too sharp.

Because the static angle is unknown, the first mode was also computed for water at $30$--$60^\circ$ and oil at $15$--$60^\circ$. This range shifts the frequency by more than the frequency errors, so frequency alone cannot test the curvature correction sharply, but it does separate the contact-line classes. In orbit, a mobile line for water or a pinned line for oil misses the measured frequency by at least $39\,\%$ at every angle, which confirms the experimenters' interpretation. Neither ideal contact-line condition reproduces the measured damping over these angle ranges.

With ideal contact lines, the wall Stokes layer accounts for about four fifths of the modelled viscous loss, while the model underestimates the damping ratios from the MODE fits by $38$--$87\,\%$. Recomputing the spectra with $\T$, $\K$ and $(\C_{\rm s},\Pim)$ set separately to their flat or curved values shows that full curvature raises the decay rate by $26$--$34\,\%$, almost entirely through the wall Stokes layer on the taller wetted wall and the reshaped field, while the bulk strain contributes at most $4\,\%$.

\subsubsection{Contact-line resistance}
\label{sec:mobility}

Both ideal lines dissipate nothing at the contact line, so the resistive regime between them was explored by sweeping the friction coefficient $\alpha$ of \eqref{eq:contact_coeffs} from free to pinned for every static angle.

\begin{figure}[!tb]
\centering
\includegraphics[width=0.66\textwidth]{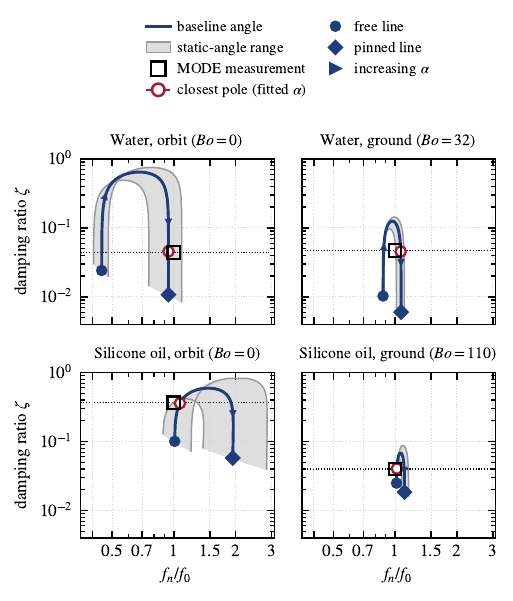}
\caption{Pole locus of the MODE fundamental as the friction coefficient $\alpha$ increases from a free to a pinned line, at the baseline static angle (line) and over the static-angle range (shading).}
\label{fig:mobility}
\end{figure}

In the resulting pole locus (Fig.~\ref{fig:mobility}) the pole moves from the free-line to the pinned value along an arch. The frequency rises, while the damping first grows and then falls, because a nearly free and a nearly pinned line both dissipate little \citep{hocking1987}. The measured pole lies inside the static-angle band in orbit and $1$--$2\,\%$ below it in frequency on the ground. At the baseline angles the best $\alpha$ reproduces it within about $7\,\%$ in frequency and $3.4\,\%$ in damping, with $\alpha=1.1\times10^4$ and $3.4\times10^3$ for water and $42$ and $17$ for oil, in orbit and on the ground. Water thus needs a nearly pinned, highly resistive line and oil a nearly free one, as the frequencies and the experimenters indicate, and as expected from the little hysteresis of silicone oil and the tens of degrees typical of water.

The response curves give a partly independent check (Fig.~\ref{fig:spectra}, Table~\ref{tab:hysteresis}). For water, the per-case resistive line cuts the force errors about fourfold, but at the two higher orbital excitations they rise again, consistent with amplitude-dependent contact-line dissipation.

With one parameter per case and an unknown angle, this agreement shows only consistency. The fitted resistance differs between orbital and ground conditions and substantially exceeds values inferred from independent water-drop mobility measurements \citep{xiasteen2018,ludwicki2022}. Since the substrates and flow regimes differ, this comparison suggests that the fitted coefficient may absorb additional dissipation. Coulomb-type hysteresis is one candidate, with equivalent resistance $c_{\rm eq}$ increasing at lower frequency and smaller contact-line amplitude.

\FloatBarrier

\subsubsection{Contact-angle hysteresis}
\label{sec:hysteresis}

\begin{figure}[!tb]
\centering
\includegraphics[width=0.72\textwidth]{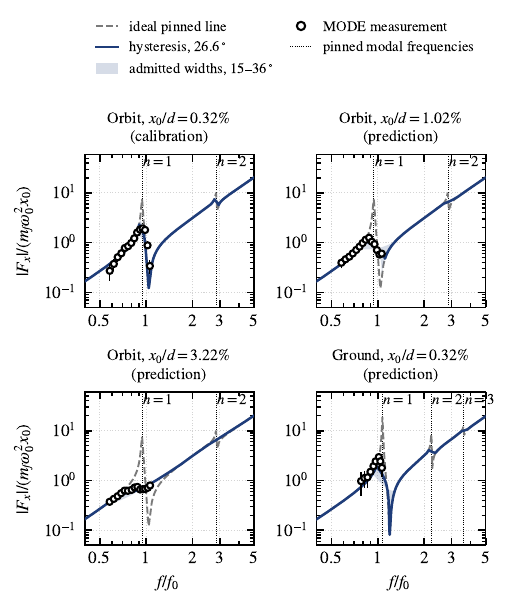}
\caption{Water force response, $|F_x|/(m_f\omega_0^2x_0)$ with $\omega_0=2\pi f_0$, for one hysteresis width, $\theta_A-\theta_R=26.6^\circ$, calibrated on the orbital response at $x_0/d=0.32\,\%$ (upper left). The other three panels are predictions. Grey dashed: ideal pinned line. Navy: hysteresis law, with shading over the calibration-admissible widths ($15$--$36^\circ$), fixed before comparison with the held-out data. Symbols: digitised MODE data. Dotted: pinned modal frequencies.}
\label{fig:hysteresis}
\end{figure}

The linear resistance is now replaced by the hysteretic part of the inclusion \eqref{eq:dvi_system}, with advancing and receding angles symmetric about $\theta_c$. The full hysteresis width is $\theta_A-\theta_R$. Each forced response is computed by harmonic balance, solving $c_{\rm eq}$ and $A_c$ together at every frequency and holding the line pinned below the breakaway threshold. The MODE dwell schedule is not documented, so frequencies are swept upwards with continuation. This width is calibrated on the orbital force at the lowest excitation ($x_0/d=0.32\,\%$) only, and the two higher orbital excitations, the orbital phase and the ground force are then predicted with no further adjustment (Fig.~\ref{fig:hysteresis}).

\begin{table}[!tb]
\centering
\caption{MODE water force NRMSE (\%) at $x_0/d=0.32\,\%$ in orbit (calibration), $1.02$ and $3.22\,\%$ in orbit, and $0.32\,\%$ on the ground. Phase RMSE uses the journal data at $0.32\,\%$. Softening is the peak shift from $0.32$ to $1.02\,\%$ in orbit (measured $-7.3\,\%$). Per-case resistance is fitted to each pole. Other fitted laws use the calibration force curve and one parameter, except the joint fit ($\theta_c=47.5^\circ$ and the width). HB: harmonic balance. Exact: periodic orbits. --: not computed.}
\label{tab:hysteresis}
\small
\setlength{\tabcolsep}{4pt}
\begin{adjustbox}{max width=\textwidth}\begin{tabular}{@{}lllrrrrrr@{}}
\toprule
Law & Added damping & Constant & cal. & $1.02\,\%$ & $3.22\,\%$ & ground & phase & softening\\
\midrule
Ideal pinned line & -- & -- & $112$ & $149$ & $208$ & $192$ & $57^\circ$ & --\\
Linear resistance, per case & $c_\ell=R\alpha\nu\sin\theta_c$ & $\alpha$ per case & $27$ & $41$ & $81$ & $28$ & $32^\circ$ & --\\
\addlinespace
Hysteresis, HB & $c_{\rm eq}=4F_h/(\pi\omega A_c)$ with stick & $\theta_A-\theta_R=26.6^\circ$ & $30$ & $13$ & $11$ & $35$ & $57^\circ$ & $-8.5\,\%$\\
Hysteresis, exact & inclusion \eqref{eq:dvi_system} & $\theta_A-\theta_R=26.6^\circ$ & $33$ & $14$ & $12$ & -- & $55^\circ$ & --\\
Hysteresis, joint & $c_{\rm eq}=4F_h/(\pi\omega A_c)$ with stick & $\theta_A-\theta_R=25^\circ$ & $13$ & $16$ & $11$ & $36$ & $38^\circ$ & --\\
\addlinespace
Line, linear & $c_{\rm eq}=R\alpha\nu\sin\theta_c$ & $\alpha=1.19\times10^4$ & $27$ & $47$ & $88$ & $88$ & $33^\circ$ & $0.0\,\%$\\
Line, quadratic & $c_{\rm eq}=8C_q\omega A_c/(3\pi)$ & $C_q=281R^2$ & $27$ & $84$ & $163$ & $141$ & $33^\circ$ & $+0.3\,\%$\\
Modal, linear & $c_m$ constant & $c_m=1.32\,$s$^{-1}$ & $25$ & $45$ & $86$ & $71$ & $30^\circ$ & $0.0\,\%$\\
Modal, quadratic & $c_m=C_m\omega A_s$ & $C_m=0.043$ & $26$ & $44$ & $86$ & $27$ & $31^\circ$ & $0.0\,\%$\\
\bottomrule
\end{tabular}\end{adjustbox}
\end{table}

For water, the calibrated hysteresis width is $26.6^\circ$, consistent with values commonly reported on engineering surfaces. Using this width, the model predicts the remaining force responses with errors of $11$--$13\,\%$ in orbit and $35\,\%$ on the ground, compared with $149$--$208\,\%$ for an ideal pinned line (Table~\ref{tab:hysteresis}). The calibration also admits widths of about $15$--$36^\circ$ within the data-extraction tolerance. Across this range, prediction errors are $7$--$30\,\%$ in orbit and $24$--$57\,\%$ on the ground, at least three times smaller than for the pinned line. The resulting response band contains $64$--$100\,\%$ of the measured points on each prediction curve (Fig.~\ref{fig:hysteresis}).

At low excitation, the contact line remains pinned except near resonance. Stronger excitation broadens the frequency range over which it slips, increasing dissipation and flattening the resonance until no distinct peak remains, in agreement with the measurements.

Between the two lowest orbital excitations, the measured force peak shifts down in frequency by $7.3\,\%$, compared with $8.5\,\%$ predicted. The ground peak is predicted within $4\,\%$. In the model, this softening comes from contact-line hysteresis, consistent with free-decay studies using the same law \citep{cocciaro1993,violagallaire2018prf}. The calculated peak surface amplitude is about $0.1R$ and changes by only $2\,\%$ between the two excitations. Nonlinear fluid effects remain possible because surface amplitudes were not measured. Although the neglected capillary energy is below $2\,\%$, this estimate does not bound the omitted nonlinear kinetic terms.

\begin{figure}[!tb]
\centering
\includegraphics[width=0.78\textwidth]{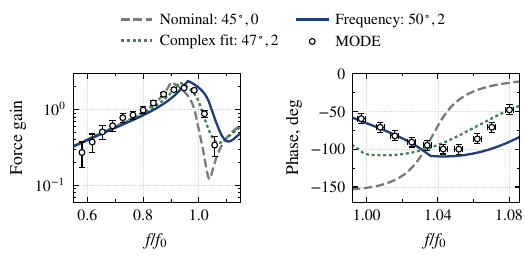}
\caption{MODE orbital water at $x_0/d=0.32\,\%$. Here $f_0=3.31\,$Hz and gain is $|F_x|/(m_f\omega_0^2x_0)$. Pairs are $(\theta_c,\chi)$ at fixed $\theta_A-\theta_R=26.6^\circ$ and $\alpha=0$. Bars show extraction tolerances. The ten NASA MODE phase points give reaction force phase relative to tank displacement \citep{mode1993}. In angle order, phase RMSEs are $57.8$, $22.1$ and $18.0^\circ$. Pointwise-relative complex RMS errors are $90.8$, $43.0$ and $97.5\,\%$ at the seven phase points with interpolated force data. At $47^\circ$, held-out orbital force errors are $16.5$ and $10.2\,\%$ (nominal $13.0$ and $10.6\,\%$), with $-9.3\,\%$ softening (measured $-7.3\,\%$). At $50^\circ$, pinned $f_n=3.295\,$Hz and the intermediate force error is $23.4\,\%$.}
\label{fig:mode_sensitivity}
\end{figure}

With the calibrated hysteresis width fixed, the equilibrium and ROM are recomputed at each static angle. Additional surface loss enters through $\D_{\rm ext}=\D+\chi\beta\Pim_s$, where $\Pim_s$ is the surface tangential-velocity Gram and $\chi$ scales the added damping. The term is purely dissipative and linear in velocity. At $\chi=1$, the inextensible-surface Stokes scale \citep{henderson1998} accounts for about a third of the orbital water damping deficit. At $(\theta_c,\chi)=(47^\circ,2)$, the phase RMSE falls from $58^\circ$ to $22^\circ$ and the complex-response error from $91$ to $43\,\%$. At $(50^\circ,2)$, the pinned frequency is closer to the MODE value and the phase RMSE falls to $18^\circ$, while the complex-response error rises to $98\,\%$ (Fig.~\ref{fig:mode_sensitivity}). The remaining differences in phase and preferred hysteresis width may arise from uncertain wetting conditions, additional surface losses, or contact-line motion that the model does not resolve. Possible mechanisms include resistance that varies with line speed and slipping confined to part of the circumference. The present comparisons cannot distinguish these effects. A nonlinear contact-line resistance, for example one based on the Cox--Voinov relation \citep{cox1986,snoeijer2013}, could be introduced through a convex dissipation potential while preserving passivity. Contact-line mobility can be measured independently on vibrated drops \citep{ludwicki2022}.

Sweeping the hysteresis width over $0$--$80^\circ$ reveals distinct force-error minima across the four MODE water conditions ($26$--$44^\circ$), while calibration-admissible widths ($15$--$36^\circ$) retain substantially lower errors than the ideal pinned-line baseline. The upward drift of the optimum with excitation and gravity may reflect rate-dependent dissipation, partial depinning or other model discrepancy, and it is reported without an additional parameter to absorb it.

With the baseline viscous model ($\chi=0$), jointly calibrating the static contact angle and hysteresis width gives $\theta_c=47.5^\circ$ and $\theta_A-\theta_R=25^\circ$. The calibration force error falls to $13\,\%$ and the predicted orbital phase error to $38^\circ$, while errors for the two higher orbital excitations and the ground response remain similar (Table~\ref{tab:hysteresis}). The static angle is more tightly constrained by this calibration than the hysteresis width.

Separate sensitivity checks assess transfer between conditions. Calibrating on each of the four water curves in turn predicts the other three with errors below $44\,\%$. Tested changes in fill and residual acceleration alter these errors by at most two percentage points. Across the calibration-admissible static angles of $46$--$50^\circ$, force errors are $9$--$21\,\%$ in orbit and $31$--$43\,\%$ on the ground, with resonance softening retained.

Four alternative one-constant laws were calibrated on the same curve and then held fixed (Table~\ref{tab:hysteresis}). A linear and a quadratic contact-line drag add $c_{\rm eq}\e\e\tp$ to $\D$, with $A_c$ the amplitude of $\e\tp\q$ and the quadratic drag $C_q|u|u$, and a linear and a quadratic modal damping add $c_m\T$ to a pinned line, with $A_s=|\hat F_x/(-m_f\omega^2\hat x_0)-1|$. All enter as first-harmonic describing functions, with constants per unit $\rho\pi$. The four laws fit the calibration curve equally well but miss the softening, and their orbital held-out errors are three to fifteen times larger. On the ground the quadratic modal law does better ($27$ against $35\,\%$), so hysteresis is not superior in every case. A resistance proportional to $(\omega A_c)^p$ predicts the higher orbital excitations within $15\,\%$ only for $-1\le p\lesssim-0.65$, a range that includes Coulomb friction and excludes linear and quadratic drag. The MODE data therefore favour a contact-line law whose effective resistance falls with amplitude, as hysteresis does, without identifying it uniquely.

For silicone oil the calibration selects an essentially free line, so the oil force data are insensitive to the contact line. First-mode fits to the digitised curves bound the orbital oil damping ratio only loosely, and the admissible range contains the prediction ($\zeta\approx0.10$), so the oil damping deficit rests on the published fit ($\zeta=0.363$) alone.

As a check on harmonic balance, exact periodic orbits of the inclusion were computed by Newton shooting on the event-resolved period map. They are all stable and change the force errors by at most $2.6$ percentage points and the phase error by about $3^\circ$ (Table~\ref{tab:hysteresis}, ``exact'').

\FloatBarrier

\subsection{Effect of gravity on the contact-line pole range}
\label{sec:gravity}
\begin{figure}[!tb]
\centering
\includegraphics[width=0.88\textwidth]{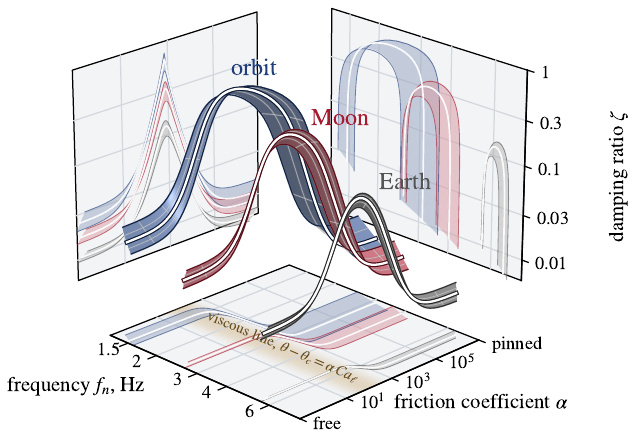}
\caption{First-mode pole of the MODE water tank in orbit (navy), on the Moon (crimson) and on the ground (grey) against the friction coefficient $\alpha$ and the static angle ($30$--$60^\circ$, white line $45^\circ$). The back plane shows frequency against damping, as in Fig.~\ref{fig:mobility}. The shaded floor band marks $\alpha\approx10$--$10^3$, expected for a viscous contact line. The lunar surface is a model prediction.}
\label{fig:locus3d}
\end{figure}

Figure~\ref{fig:locus3d} maps the MODE water pole locus versus static angle and $\alpha$ in orbit, on the Moon and on the ground (axial $\Bo=0$, $5.3$ and $32$). The frequency switches from free to pinned within less than a decade of $\alpha$ around the damping peak at $\alpha\approx750$. Increasing acceleration from orbit through lunar to terrestrial gravity raises the frequencies and reduces the contact-line pole range. From orbit to ground, the baseline pinned-to-free frequency ratio falls from $2.1$ to $1.2$, while maximum damping falls about fivefold but remains above the ideal-line values. At fixed contact-line amplitude and modal factors, $c_{\rm eq}\propto1/\omega$ gives a hysteretic damping-ratio contribution proportional to $1/\omega^2$, consistent with axisymmetric friction damping \citep{utsumi2017}.

Gravity weakens the contact-line influence on the fundamental pole, so ground tests capture only part of the orbital uncertainty. Since $\Bo\propto R^2$, a $0.5\,$m water tank already has $\Bo>5000$ on the Moon. Curvature is confined to the wall capillary length (about $7\,$mm), and the arch approaches the flat-interface pole.


\FloatBarrier

\section{Conclusions}
\label{sec:conclusions}

The port-Hamiltonian formulation integrates curved-meniscus inertia, viscous losses and nonsmooth contact-line hysteresis within one passive dynamical system. Analytical stick--slip propagation supports faster-than-real-time coupled simulation. The energy and dissipation structure also provides a basis for future membrane and mechanical coupling through energy exchange ports, and extension to other tank geometries and topologies.

The model agrees with an independent Stokes calculation of viscous decay within $5\,\%$ and reproduces the Henderson--Miles modes and Cocciaro hysteretic decay. Curvature correction reduces the RMS frequency error across $42$ NASA drop-tower and ground measurements from $24$ to $4\,\%$, while MODE frequencies agree within the uncertainty of the unknown wetting conditions.

For MODE water, one hysteresis width calibrated on the lowest orbital excitation predicts the two higher excitations with force errors of $11$--$13\,\%$ and the ground response with $35\,\%$, compared with $149$--$208\,\%$ for an ideal pinned line. It also predicts $8.5\,\%$ resonance softening against $7.3\,\%$ measured, capturing an amplitude dependence missed by the tested one-constant linear and quadratic damping laws. These held-out comparisons support the model's predictive transfer within MODE, although they do not uniquely identify the dissipation mechanism.

Contact-line conditions have their strongest influence in orbit, where the pinned-to-free frequency ratio is $2.1$, compared with $1.2$ on the ground. Until the wetting and contact-line behaviour of flight hardware are characterised, the remaining response uncertainty calls for damping ranges in GNC analyses.

\section*{Declaration of competing interest}
The authors declare no known competing financial interests or personal relationships that could have appeared to influence the work reported in this paper.

\section*{Funding}
This work was supported by the Italian Ministry of University and Research (MUR) through a PNRR innovative doctoral fellowship co-financed by Thales Alenia Space Italia under the European Union NextGenerationEU programme.

\section*{Data availability}
The data and code supporting the findings of this study are available from the corresponding author upon reasonable request.

\section*{CRediT authorship contribution statement}
\textbf{Gianni Cassoni:} Conceptualization, Methodology, Software, Validation, Formal analysis, Investigation, Data curation, Visualization, Writing -- original draft, Writing -- review \& editing.
\textbf{Fabrizio Gennari:} Supervision, Funding acquisition.
\textbf{Jacopo Serafini:} Supervision, Funding acquisition.
\textbf{Pierangelo Masarati:} Supervision, Funding acquisition.

\section*{Declaration of generative AI and AI-assisted technologies in the manuscript preparation process}
The authors used Claude (Anthropic) and ChatGPT/Codex (OpenAI) for language editing and analysis assistance. They reviewed the code, results and interpretation and take full responsibility for the publication.

\bibliographystyle{elsarticle-num-names}
\bibliography{references}

\end{document}